\documentstyle[aps,eqsecnum,epsfig]{revtex}

\begin{document}

\draft

\title{Two-point correlation function in systems with van der Waals type
  interaction}

\author{Daniel Dantchev}

\address{Institut f\"{u}r Theoretische Physik, Technische Hochschule
  Aachen, 52056 Aachen, Germany\\ and\\
Institute of Mechanics, Bulgarian Academy of Sciences, Acad.
G. Bonchev St.  Building 4, 1113 Sofia, Bulgaria}

 \date{\today}

 \maketitle
\begin{abstract}
The behavior of the bulk two-point correlation function $G({\bf
r};T|d)$ in $d$-dimensional system with van der Waals type
interactions is investigated  and its consequences on the
finite-size scaling properties of the susceptibility in such
finite systems with periodic boundary conditions is discussed
within mean-spherical model which is an example of Ornstein and
Zernike type theory. The interaction is supposed to decay at large
distances $r$ as $r^{-(d+\sigma)}$, with $2<d<4$, $2<\sigma<4$ and
$d+\sigma \le 6$.  It is shown that $G({\bf r};T|d)$ decays as
$r^{-(d-2)}$ for $1\ll r\ll \xi$, exponentially for $\xi\ll r \ll
r^*$, where $r^*=(\sigma-2)\xi \ln \xi$, and again in a power law
as $r^{-(d+\sigma)}$ for $r\gg r^*$. The analytical form of the
leading-order scaling function of $G({\bf r};T|d)$ in any of these
regimes is derived.

\end{abstract}


\section{Introduction}
\label{intro} It is well known that the critical properties of a
given statistical-mechanical system depend only on a small number
of parameters like the dimensionality $d$ of the system, the
symmetry of the order parameter characterizing the corresponding
phase transition, and on general properties of the interaction
coupling the order parameter at different locations. For example,
in an isotropic $O(n)$ system one expects that all critical
exponents and scaling function of a given physical quantity are
independent on, say, lattice structure, or on short-range details
of the interaction. Let us, for definiteness of notation, speak
about Ising-like systems (i.e. $n=1$) with Hamiltonian
\begin{equation}
{\cal H}= -\frac{1}{2} \sum_{{\bf r}\ne{\bf r}'} J({\bf
  r}-{\bf r}')S_{\bf
  r}S_{{\bf r}'}.
\label{ham}
\end{equation}
In the context of the critical phenomena the usual criterion for a given
interaction to be considered as short-ranged is finite second moment of
$J({\bf r})$, i.e. in terms of the Fourier transform $\tilde{J}({\bf
    k})$ of this interaction for small $k=|{\bf k}|$ one has
\begin{equation}
\tilde{J}({\bf k})=\tilde{J}_0+\tilde{J}_2 k^2+\Delta\tilde{J}({\bf
  k}),
\label{int}
\end{equation}
where $\Delta\tilde{J}({\bf k})$ is asymptotically smaller than
$k^2$. Then, if $2<d<4$, for the bulk two-point correlation function
\begin{equation}
  G({\bf r};T|d)=:<S_{\bf r}S_{\bf 0}>-<S_{\bf 0}>^2
\end{equation}
one normally writes
\begin{equation}
G({\bf r};T|d)=D r^{-(d-2+\eta)} X^{\pm}({\bf r}/\xi),
\label{cor}
\end{equation}
where $X^{\pm}$ are two universal scaling functions (for $T>T_c$ and
$T<T_c$, respectively), $D=D(T)$ is a nonuniversal, slowly varying
function of temperature that well can be approximated by a constant in
the vicinity of the critical point $T=T_c$ and $\xi$ is the bulk
correlation length, i.e. $\xi(T)\simeq \xi_0^+ t^{-\nu}$,
$t\rightarrow 0^+$ with $t=(T-T_c)/T_c$. For $T \ge T_c$ one has
\begin{equation}
X^{+}(x) \simeq \left\{
\begin{array}{lc}
\hat{X}^+ x^{(d-3)/2+\eta}\exp(-x),
&x\rightarrow\infty \\
& \\
\mbox{constant}, & x\rightarrow 0.
\end{array}
\right.
\label{cplus}
\end{equation}
The above is, in fact, the classical result of Fisher \cite{F64}
for the two-point correlation function in the critical region of a
simple fluid. Note, that when $r\gg\xi$ the correlations decay
exponentially fast with the distance $r$. It is well known that in
simple nonpolar fluids the interactions are characterized by
potentials that decay as inverse powers of the distance at large
$r$. In $d=3$ the most prominent case is the induced
dipole-induced dipole (or van der Waals) interaction for which
(neglecting the retardation effects) the potential decays as
$\Phi({\bf r}):=-J({\bf r})=-A/r^6$, where $A>0$ is a positive
constant. One easily can check that this interaction has a Fourier
transform which is indeed of the type given in Eq. (\ref{int}).
But for such systems ($d$-dimensional Ising model), in which the
interactions decrease  in inverse power with respect to the
distance between the interacting objects, i.e. as
$r^{-(d+\sigma)}$, $\sigma>2$, the following {\em rigorous} result
due to Iagolnitzer and Souillard \cite{IS77} is available.

{\it Theorem}\cite{IS77}: The two-point correlation function $G({\bf
  r};T,H|d)$ of a ferromagnetic system in the presence of an external
magnetic field $H$ does not decay faster than its potential $J$.
For any $T<\infty$ and any real $H$ there exists a strictly
positive constant $C(T,H)$, such, that
\begin{equation}
G({\bf r};T,H|d)>C(T,H)J({\bf r}). \label{theorem}
\end{equation}

For $H=0$ the theorem is valid for $T>T_c$ and for any of the
two ``pure'' phases (the ``plus'' and the ``minus'' ones) for $T<T_c$.

The immediate consequence of this theorem is that if $T \ne T_c$
(\ref{cor}) - (\ref{cplus}) could not be true for $r$ large enough
independently on how close $T$ is to the critical point. The only
way to reconcile (\ref{cor}) - (\ref{cplus}) with the above
theorem is to realise that if $T \ne T_c$ (\ref{cor}) -
(\ref{cplus}) could be valid only up to some $r=r^*(T)$. Then for
$1 \ll r \ll \xi$ the correlations will decay as
$r^{-(d-2+\eta)}$, for $\xi \ll r \ll r^*$ they will fall off
exponentially, but, for $r>r^*$ they should again decay in a power
law as a function of the distance, namely as $r^{-(d+\sigma)}$. In
other words one should observe a crossover from power law to
exponential and then, again, to power law behavior of the
correlations. Saying this one immediately stacks with at least two
important questions that appear naturally: 1) What is the value of
$r^*$, i.e. where this crossover happens and 2) What are the
properties of the function describing that crossover.  One of the
aims of the current article is to answer those questions in the
framework of an exactly solvable model.

It is easy to check that one has the above situation only with
interactions of the type $J({\bf r})\simeq A/r^{d+\sigma}$, where
$\sigma>2$. To avoid misunderstanding in the remainder of the text
let us make the following {\it definitions}. {\it i)} An
interaction will be called of short range if for any finite $m$
its $m$-th moment is finite, i.e. $\sum_{\bf r} r^m J({\bf
r})<\infty$. {\it ii)} An interaction is long ranged if there
exists a finite $m$ such that the corresponding $m$-th moment
diverges. If $m=2$ this is a leading-order long range interaction,
and if $m>2$ this is a subleading (van der Waals type) long-range
interaction.

We recall that even if the interaction is short-ranged in the
above sense (this is the situation we have with nearest neighbour,
next-nearest neighbour, etc. interactions, i.e. with interactions
that are essentially of a finite range), then (\ref{cor}) -
(\ref{cplus}) are again valid only for $1<<r<<r^*_{{\rm sr}}$. The
exact results for $d=2$ Ising model (see, e.g. McCoy and Wu
\cite{MW73}) and the mean-field results of Fisher and Burford
\cite{FB67} suggest  that, if $T\ne T_c$, $r^*_{{\rm sr}}\sim
\xi^2$. (The exact calculations due to Chen and Dohm \cite{CD2000}
for the spherical model give a bit more ``generous'' estimation
for $r^*_{{\rm sr}}$, namely $r^*_{{\rm sr}}\sim \xi^3$, see also
below.) For $r>>r^*_{{\rm sr}}$ the interactions decay, of course,
again exponentially, but they contain a nouniversal prefactor
\cite{CD2000}, i.e. their {\it leading-order} behavior is then
nonuniversal. If the interaction is of a leading long-range type
then (\ref{cor}) is also valid but in the limit $r/\xi\gg 1$ one
has to require that  $X^{+}(x) \simeq \hat{X}^+
x^{\eta-2-\sigma}=\hat{X}^+ x^{-2 \sigma}$, where we have taken
into account that $\eta=2-\sigma$ if $\sigma <2$
\cite{FMN72,M73,BJG76}. The corrections to the large distance
correlations in this case are in a  power-law of $r$, which means
that their leading order behavior is universal for any $r>>1$.
This asymptotic is confirmed by the exact results for the
spherical model due to Joyce \cite{J66,J72} and it is in tune with
the above theorem for the Ising model. Note that for $r$ large
enough the correlations always fall off in a power law with the
distance with the only exception of  interactions of a fully
finite range when they do decay exponentially.

That (\ref{cor}) - (\ref{cplus}) should be modified for the case of
subleading long-range interactions has been noticed by several authors.

First Widom proposed \cite{W64} that for $r\rightarrow\infty$
\begin{equation}
G({\bf r};T|d)\simeq \beta J({\bf r}) + a_1 r^{-(d-1)/2}\exp(-r/a_2),
\end{equation}
where $\beta=1/(k_B T)$, $k_B$ being the Boltzmann's constant, and
$a_1$ and $a_2$ are "depending only on the thermodynamics state
constants". It is clear that in nowadays formulations the above
means to take for the correlation function a sum of $\beta J({\bf
r})$ and the right-hand side of Eq. (\ref{cor}). Later the problem
has been attacked by Enderby, Gaskell and March \cite{EGM65}. They
consider a three-dimensional fluid, i.e. the case $d=\sigma=3$.
Supposing the Ornstein-Zernike integral equation to be valid and
taking the direct correlation function to be $c(r)=\beta J(r)$,
they obtain, after assuming that the structure factor $S$ has a
Fourier transform of the type $S(k)=\chi/\beta + c_2 k^2 + c_3 k^3
+ \cdots$, that $G({\bf r};T|3) \simeq J(r)
\chi^2/\beta=A\chi^2/(\beta r^6)$, when $r\rightarrow\infty$ and
in temperature regions "well away" from the critical point
\cite{note2}. Starting from this result, Kayser and Ravech\'{e}
\cite{KR84} suggest that $G({\bf r};T|d)$ can be decomposed in two
additive contributions $G^{{\rm sr}}$ and $G^{{\rm lr}}$, where
$G^{{\rm sr}}$ is given by (\ref{cor}), plus higher-order terms
that account for the usual corrections to scaling, and $G^{{\rm
lr}}=\Theta(r-r^*) J(r) \chi^2/\beta$. Here $\Theta(x)$ is the
Heaviside step function and $r^*$ is to be determined by the
requirement that at this point $G^{{\rm
    sr}}=G^{{\rm lr}}$. Taking $\chi\propto\xi^{2-\eta}$ in the
expression for $G^{{\rm lr}}$, which in fact means supposing $G({\bf
  r};T|d) \simeq J(r) \chi^2/\beta$, $r\rightarrow\infty$, to be valid for
general $d$ and also for temperatures close to $T_c$, the above
authors derived
\begin{equation}
r^*=(\sigma-2+\eta)\xi \ln \xi
\label{rhoRK}
\end{equation}
In \cite{FD95}  Fl\"{o}ter and Dietrich make similar statements for
$r^*$ \cite{note} for the case $d=\sigma=3$.

In the present article we will investigate the large $r$ behavior of
the correlations and will derive the explicit form of $G^{{\rm
    lr}}$ within the mean spherical model. The
interaction will be supposed to be of the type $J=A/r^{d+\sigma}$,
with $2<d<4$, $2<\sigma<4$ and $d+\sigma\le 6$.

If one knows $G$ one can immediately determine the behavior of the
bulk susceptibility $\chi$ by using the fluctuation-dissipation
relationship $\chi(T|d)=\beta \sum_{\bf r} G({\bf r};T|d)$.
Definitely, if the finite-size two-point correlation function is
known for a given {\it finite} system with a characteristic size
$L$, then one can determine in this way also the behavior of the
finite-size susceptibility $\chi(T;L|d)$.  In a recent article
\cite{CD2000} Chen and Dohm have addressed the question: could one
say what should be the scaling structure of the {\it finite-size}
susceptibility under periodic boundary conditions if one knows
only the {\it bulk} two-point correlation function? They suggest a
hypothesis, that this is possible by interpreting in a proper way
the functional dependence of $G$ on $r$ as a dependence of
$\chi(T;L|d)$ on $L$. In the present article we check the
relationship that they suggest between $G$ and  $\chi(T;L|d)$ on
the example of our exactly solvable model.  For that aim we will
use the results for $\chi(T;L|d)$ derived in \cite{DR2001} for the
same model.

The structure of the article is as follows. In Section \ref{model}
we briefly describe the model and present our starting analytical
expressions. Section \ref{ldb} contains our results for the large
$r$ behavior of the two-point correlation function $G$. In
Appendix \ref{A} we present some details of the calculations
needed to determine the asymptotics of $G$ as a function of $r$
and $\xi$. Section \ref{fsss} comments on the relationship between
the derived results for $G$ and the behavior of the finite-size
susceptibility of systems with subleading long-range interactions.
The article closes with a Discussion (Section  \ref{d}) where we
speculate about the possible extensions of our results for other
models.

\section{The model}
\label{model}

We consider a $d$-dimensional mean spherical model \cite{BC52},
\cite{LW52} (for a comprehensive review on the results available
for this model see \cite{BDT00}). The degrees of freedom consist
of a set of $N$ localised spins with Gaussian weight, and the
Hamiltonian is given by Eq. (\ref{ham}). The interaction $J({\bf
r})$ is supposed to be of van der Waals type, i.e. its Fourier
transform is supposed to be of the form
\begin{equation}
  \tilde{J}({\bf k})\simeq {\tilde J}({\bf 0})
\left(1-v_2 k^2+v_\sigma k^\sigma-v_4 k^4
    +
O(k^6)\right),
\label{int2}
\end{equation}
where $k=|{\bf k}|$, $4>\sigma>2$ and $\tilde{J}({\bf 0})$,
$v_2$, $v_\sigma$ and $v_4$
are nonuniversal positive constants.
 Note that the signs
of the coefficients in the small $k$ expansion of the Fourier
transform of the interaction are chosen so as they normally appear
for subleading long-range interactions that decay in power law
with the distance between the interacting objects - molecules or
spins. In (\ref{int2}) $\tilde{J}({\bf 0})$, $v_2$, $v_\sigma$ and
$v_4$ are $\sigma$-dependent --- for simplicity of notation this
dependence is omitted here. The term $v_{\sigma} q^{\sigma}$ in
(\ref{int2}) is associated with a contribution to the real-space
interaction going as $r^{-d-\sigma}$. Furthermore, we suppose that
$\tilde{J}({\bf k})- \tilde {J}({\bf 0})<0$    if ${\bf k}\ne {\bf
0}$, which reflects the fact that there are no competing
interactions in the system and that the only ground state is the
ferromagnetic one.  Of course, it would be interesting to consider
such systems
--- say with a combination between antiferromagnetic short range and
ferromagnetic subleading long-range interactions, but this is out
of the scope of the current article.

The partition function of the model is given by the multiple
integral
\begin{equation}
\int_{-\infty}^{\infty} ds_{1} \cdots \int_{-\infty}^{\infty}ds_{N}
\exp \left[ -\beta {\cal H} \right],
        \label{partfun1}
\end{equation}
supplemented by the mean spherical condition
\begin{equation}
\sum_{i=1}^{N} \langle s_{i}^{2}\rangle =N,
        \label{msc1}
\end{equation}
which can be enforced with the use of a ``Lagrange multiplier'' term
going as $ \lambda \sum_{i=1}^{N}s_{i}^{2}$ into the effective
Hamiltonian, and thence into the partition function.  The spherical
model equation of state then takes the form
\begin{equation}
\sum_{{\bf k}}\frac{k_{B}T}{ \lambda-\tilde {J}({\bf 0})(1 - v_{2}k^{2} +
v_\sigma k^{\sigma}- v_4 k^4)} = N.
        \label{msc2}
\end{equation}
The phase transition in this model occurs when the combination
$\lambda-\tilde{J}$ takes on a value asymptotically close to zero.
The difference between the equation of state in (\ref{msc2}) and
the standard mean spherical model condition in short range systems
lies in the addition of the term going as $k^{\sigma}$ in the
denominator on the left hand side of (\ref{msc2}).  In general,
this term is taken to be negligible, but we will soon see that it
leads to interesting effects.

For the model defined in the above way  it can be shown, following \cite{J72},
that the bulk correlation function $G({\bf r};K|d,\sigma)$
is given by, if $2<d<4$,
\begin{equation}
  G({\bf r};K|d,\sigma)=\frac{1}{K}\frac{1}{(2\pi)^d}\int_{{\cal R}^d}
\frac{e^{i{\bf k}.{\bf r} }d{\bf r}}{\xi^{-2}+k^2-b k^\sigma+ c k^4},
\label{cord}
\end{equation}
where $K=\beta v_2 \tilde{J}({\bf 0})$, and $b=v_\sigma/v_2>0$ and
$c=v_4/v_2>0$
are nonuniversal constants. Let us note that the values of $b$ and $c$
are such, that there are no real roots of the equation
$1-bk^{\sigma-2}+ck^2=0$. The last follows from the propositions we
made for $\tilde{J}({\bf k})$.  Note also that in (\ref{cord}) we have
taken the
cut-off in the $k$-space to be infinity (for a lattice system it will
mean that one considers the limit of a zero lattice spacing).
This is possible because of the rapid oscillations of the
exponential function in the integrand, but in this way we neglect all
finite cut-off effects that will give nonuniversal contributions
towards the critical behavior of the two-point correlation function
(see \cite{CD2000} for details).
In (\ref{cord}) $\xi=\xi_2$ is the second moment correlation length
defined via (see, e.g. \cite{A84})
\begin{equation}
\xi^2=-\left[\tilde{G}({\bf 0};K|d,\sigma)\right]^{-1}
\left.\frac{\partial}{\partial k^2}
\tilde{G}({\bf k};K|d,\sigma)\right |_{k=0},
\end{equation}
where $\tilde{G}({\bf k};K|d,\sigma)$ is the Fourier transform of
$G({\bf r};K|d,\sigma)$. Because of this identification one can, in
fact, skip for our purposes the analysis of the spherical field
equation (\ref{msc2}) - one directly has
$\lambda=\tilde{J}({\bf 0})(1+v_2 \xi^{-2})$.

Since $\sigma>2$ and since we are interested in the behavior of
$G({\bf r};K|d,\sigma)$  for $|{\bf r}|\gg 1$ (note that then the
leading order contributions of the integral in (\ref{cord}) will
be coming from  small $k$ values), one can rewrite (\ref{cord}) in
the form
\begin{equation}
G(r;K|d,\sigma)=G^{\rm{sr}}(r;K|d)+G^{\rm{lr}}(r;K|d,\sigma),
\label{cs}
\end{equation}
i.e. as a sum of ``short-range'' and ``long-range'' parts. The ``short-range''
correlation function is the part that is only due to the short-range
component of the
interaction and,  as usual, will be taken to be of the form
\begin{equation}
G^{\rm{sr}}(r;K|d)=\frac{1}{K}\frac{1}{(2\pi)^d}\int_{{\cal R}^d}
\frac{e^{i{\bf k}.{\bf r}} d{\bf r}}{\xi^{-2}+k^2}.
\label{csr}
\end{equation}
The other contributions that are due to the subleading
components of the interaction do form the corresponding ``long-range''
part. As it has been already stated in Section \ref{intro} such
a structure has been supposed to hold by Kayser and Ravech\`e \cite{KR84}
in their qualitative analysis of the correlation functions in
fluids.

Performing the integrations in (\ref{csr}) and taking into account
that for
$2<\sigma<4$ and $r \gg 1 $
\begin{eqnarray}
G(r;K|d,\sigma) & = &  G^{\rm{sr}}(r;K|d) +
\frac{1}{K}\frac{1}{(2\pi)^d}\int_{{\cal R}^d}
\frac{(b k^\sigma- c k^4)e^{i{\bf k}.{\bf r} }d{\bf r}}
{(\xi^{-2}+k^2-b k^\sigma+ c k^4)(\xi^{-2}+k^2)}
\nonumber \\
& \simeq & G^{\rm{sr}}(r;K|d) +
\frac{b}{K}\frac{1}{(2\pi)^d}\int_{{\cal R}^d}
\frac{k^\sigma \ \ e^{i{\bf k}.{\bf r} }d{\bf r}}
{(\xi^{-2}+k^2)^2} +\cdots,
\end{eqnarray}
we obtain
\begin{equation}
G(r;K|d,\sigma) = \frac{1}{K}\frac{1}{(2\pi)^{d/2}}r^{-(d-2)}
\left[X^{\rm{sr}}(r/\xi)
+br^{-(\sigma-2)}X^{\rm{lr}}(r/\xi) +\cdots \right],
\label{cl}
\end{equation}
where
\begin{equation}
X^{\rm{sr}}(x)=x^{(d-2)/2}K_{(d-2)/2}(x),
\label{cse}
\end{equation}
\begin{eqnarray}
X^{\rm{lr}}(x) & = & \frac{\pi}{\sin[(d+\sigma)\pi/2]} \left\{
    2^{d/2+\sigma-4}\ \
    _1\tilde{F}_2(2;2-\sigma/2,3-d/2-\sigma/2;\frac{x^2}{4})
     \right. \nonumber \\
& & \left. - \frac{1}{4}x^{d/2+\sigma-3}\left[xI_{d/2}(x)-
(d+\sigma-2)I_{d/2-1}(x) \right] \right\},
\label{cle}
\end{eqnarray}
and $\cdots$ stays for contributions which are corrections with respect
to the terms retained.
Here $I_a(x)$ is the modified Bessel function, and $_p\tilde{F}_q({\bf
  a};{\bf b};z)$ is the regularized generalized hypergeometric
function
\begin{equation}
_p\tilde{F}_q({\bf a};{\bf b};z)= \frac{ \ _pF_q({\bf a};{\bf
b};z)}{\Gamma(b_1)\Gamma(b_2)\cdots\Gamma(b_q)}, \label{rghf}
\end{equation}
where $ _pF_q({\bf a};{\bf b};z)$ is the generalized hypergeometric
function
\begin{equation}
 _pF_q({\bf a};{\bf b};z)=
\sum_{k=0}^{\infty}
\frac{(a_1)_k(a_2)_k\cdots(a_p)_k}{(b_1)_k(b_2)_k\cdots(b_q)_k}\frac{z^k}{k!},
\label{ghf}
\end{equation}
The symbol $(a)_k=a(a+1)\cdots(a+k-1)=\Gamma(a+k)/\Gamma(a)$ in the
above equation is the
Pochhammer's symbol. The function $_p\tilde{F}_q$ is finite for all
finite values of its arguments.  In the above expressions only the
leading order long-range
contributions (i.e. the contributions "proportional to $b$"), have
been retained and we have supposed that $2<d<4$, $2<\sigma<4$,
and $d+\sigma<6$.

We recall that for the Ornstein-Zernike type theories (including
the   mean-spherical model, see, e.g. \cite{J72}) $\eta=0$.  In
Section \ref{d} we will discuss briefly the generalization of
(\ref{cl}) for models with $\eta \ne 0$.

The expressions (\ref{cl})-(\ref{cle}) are the analytical basis for
our further analysis. Let us note that the correlations within the
spherical model have been a subject of detailed investigations (see
\cite{J72} and \cite{BDT00} for a comprehensive
review) for both short-range and leading long-range interactions.
Surprisingly enough, they have  never been investigated for subleading
long-range interactions.

\section{Large distance behavior of the bulk two-point correlation
  function}
\label{ldb}

The asymptotics of the scaling function $X^{{\rm sr}}$ for $2<d<4$ are
well known (see, e.g. \cite{D96} and references cited therein)
\begin{equation}
X^{{\rm sr}}(x) \simeq \left\{
\begin{array}{lc}
\sqrt{\frac{\pi}{2}}x^{(d-3)/2}\exp(-x) (1+O(x^{-1})),
&x\rightarrow\infty \\
& \\
\frac{\Gamma(d/2-1)}{2^{(4-d)/2}}+\frac{\pi x^{d-2}}{2^{d/2}\sin(\pi
  d/2)\Gamma(d/2)}+O(x^2), & x\rightarrow 0.
\end{array}
\right.
\label{csa}
\end{equation}
Let us, nevertheless, make some comments here. First, let us note
that the above asymptotic is obtained if one makes a quadratic
approximation of the spectrum and lets the cut-off $\Lambda$ of
the theory go to infinity. (If one keeps a sharp finite cut-off
$\Lambda$ with such an approximation of the spectrum one will
obtain a nonexponential oscillatory power-law behavior
\cite{CD2000}, \cite{CDnew}). Second, on a lattice, for nearest
neighbours interactions between the spins embedded in a
$d$-dimensional cube it has been shown that the above expression
is valid \cite{CD2000} only for $1 << r << \xi^3$. If $r \ge
\xi^3$ the correlations do depend (up to the {\em leading} order)
on the mutual positions of the spins involved, i.e. the lattice
anisotropy comes into the play and can no longer be neglected
\cite{CD2000}. So, one can think that the above expression is
valid in the region $1<<x<<\xi^2$. Third, for $x \rightarrow 0$
the second term in this short-range expansion of the correlation
function involves a $t$ dependence of the type $t^\alpha$ if one
takes into account that $\xi\simeq \xi_0^+ \ t^{-\nu}$ for
$t\rightarrow 0^+$ and that $(d-2)\nu=\alpha$, i.e.
\begin{equation}
X^{{\rm sr}}(r/\xi) \simeq
\frac{\Gamma(d/2-1)}{2^{(4-d)/2}}+
\frac{\pi  (r/\xi_0^+)^{d-2}}{2^{d/2}\sin(\pi
  d/2)\Gamma(d/2)}t^\alpha +O((r/\xi)^2),  \ \xi \gg r.
\end{equation}
The temperature dependent term in this expansion is usually not
explicitly specified in the literature on the spherical model.
Finally, let us note that according to the above asymptotics and
under the approximations made $\xi_2=\xi_e=\xi$, i.e. the
second-moment correlation length coincides in such a theory with
the exponential-decay correlation length. (In \cite{CD2000} it has
been shown that for a model on a hypercubic lattice
$\xi_2=a/[2\sinh(a/2\xi_e)]$, where $a$ is the lattice spacing and
$\xi_e$ has been chosen to be along one of the principal axis of
the lattice.)

The asymptotics of $X^{{\rm lr}}$ for $2<d<4$, $2<\sigma<4$,
$d+\sigma<6$ are (see Appendix \ref{A} for a derivation)
\begin{equation}
X^{{\rm lr}}(x)\simeq \left\{
\begin{array}{lc}
-2^{\sigma+d/2-2}
\frac{\sigma(d+\sigma-2)\Gamma\left((d+\sigma)/2-1\right)}{\Gamma(1-\sigma/2)}
x^{-4}+
O(x^{-6}),& x \rightarrow \infty \\
& \\
2^{\sigma+d/2-4}\frac{\Gamma\left((d+\sigma)/2-2\right)}{\Gamma(2-\sigma/2)}-
x^{d+\sigma-4}\frac{\pi
  (d/2+\sigma/2+1)}{2^{d/2}\Gamma(d/2)\sin\left(\pi(d+\sigma)/2\right)}+
O(x^2),& x\rightarrow 0.
\end{array}
\right.
\label{cla}
\end{equation}

The above asymptotics lead to the following behavior of the short and
long-range parts of the bulk two-point correlation function
\begin{equation}
G^{{\rm sr}}(r;K|d)\simeq \frac{1}{K}
\frac{\sqrt{\pi/2}}{(2\pi)^{d/2}}\xi^{-(d-3)/2}r^{-(d-1)/2}\exp(-r/\xi)(1+
O\left((r/\xi)^{-1})\right),
r\gg\xi,
\label{cfsa}
\end{equation}
and
\begin{equation}
G^{{\rm lr}}(r;K|d,\sigma)\simeq
-\frac{b}{K}\frac{2^{\sigma-2}}{\pi^{d/2}}\sigma(d+\sigma-2)
\frac{\Gamma((d+\sigma)/2-1)}{\Gamma(1-\sigma/2)}
\xi^{4}r^{-(d+\sigma)}, r\gg\xi. \label{cfla}
\end{equation}
\begin{figure}
\epsfxsize=4.in \centerline{\epsffile{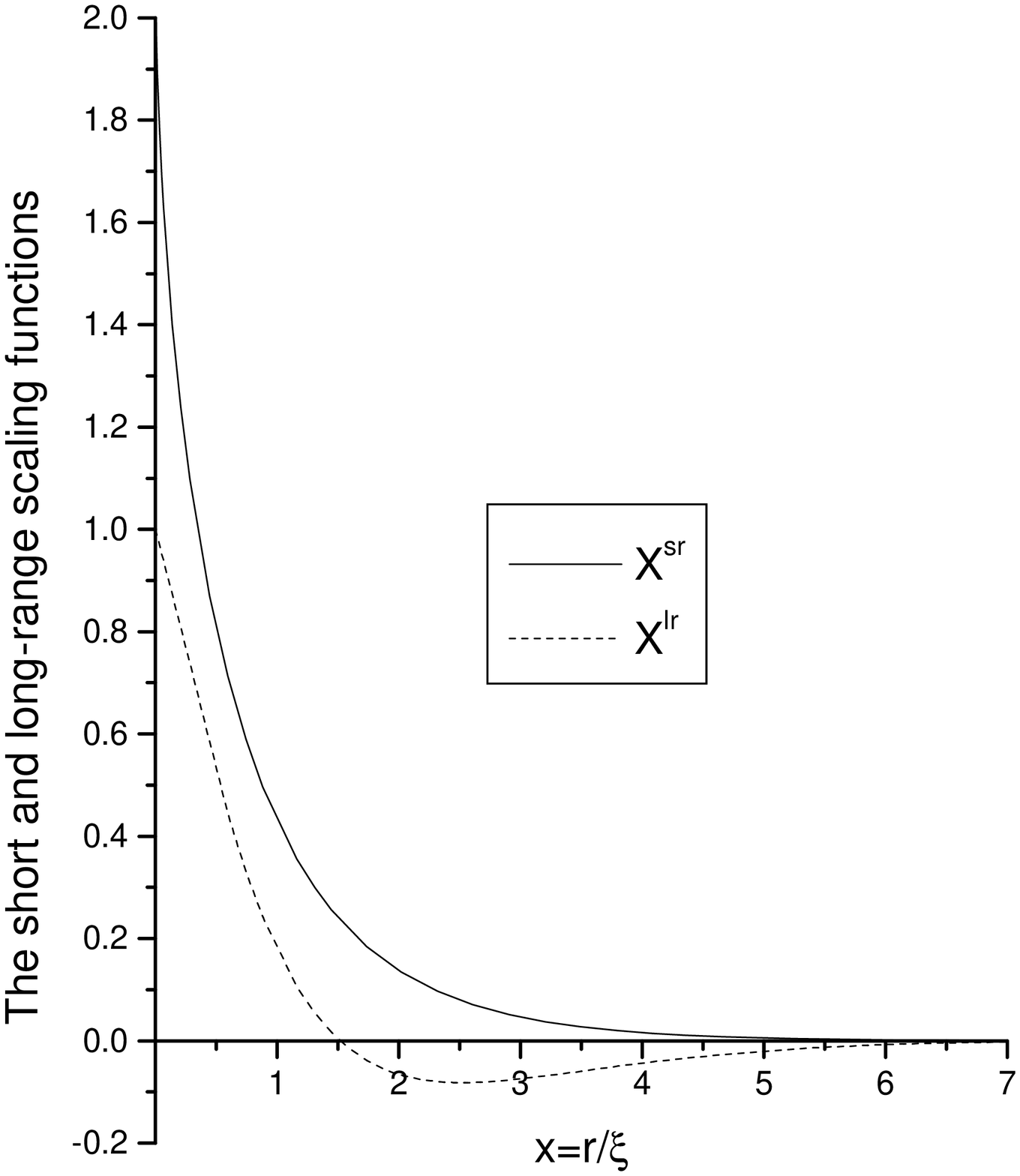}}
\vspace{0.1in} \caption{ We present as illustration  plots of the
scaling functions $X^{{\rm sr}}(x)$ and $X^{{\rm lr}}(x)$ of the
short-range and long-range correlation functions, respectively, as
a function of the scaling variable $x=r/\xi$. $X^{{\rm sr}}(x)$ is
plotted for $d=2.5$, whereas $X^{{\rm lr}}(x)$ is for  $d=3$ and
$\sigma=2.5$. For any $2<d<4$, $X^{{\rm sr}}$ is always positive
and decays monotonically as a function of $x$. For large values of
x, $X^{{\rm sr}}$ decays exponentially fast. Note that in contrast
with $X^{{\rm sr}}(x)$, $X^{{\rm lr}}(x)$ is {\it not} a monotonic
 function of $x$. In addition $X^{{\rm lr}}$ can be both
positive and negative. $X^{{\rm lr}}(x)$ decays in a power law, as
$x^{-4}$, for large values of its argument. } \label{xsr_xlr_fig}
\end{figure}

One can determine the crossover region where the correlations from
short range become long range type. To that aim one has to solve
the equation
\begin{equation}
G^{{\rm sr}}(r;K|d)\simeq G^{{\rm lr}}(r;K|d,\sigma).
\end{equation}
Having in mind Eqs. (\ref{cl}), (\ref{cfsa}) and (\ref{cfla}) one
obtains that the crossover takes place at $r\simeq r^*$, where
\begin{equation}
r^*=(\sigma-2)\xi \ln{\xi}+\left(\frac{d+1}{2}+\sigma\right)\ln\ln
\xi. \label{cover}
\end{equation}
The leading-order term of this result coincides with that
one given in \cite{KR84} (if one takes into account that $\eta=0$ for
the model under consideration).

For $d=\sigma=3$, i.e. for the true van der Waals interaction, the
corresponding scaling functions are
\begin{equation}
X^{{\rm sr}}(x) =\sqrt{\frac{\pi}{2}}\exp(-x), \label{xsr3}
\end{equation}
and
\begin{equation}
X^{{\rm lr}}(x)=\sqrt{\frac{2}{\pi}}\left\{ 1-\frac{3}{4}x \left[\exp(-x){\rm
    Ei}(x)-\exp(x){\rm Ei}(-x)\right]+ \frac{1}{4}x^2 \left[\exp(-x){\rm
    Ei}(x)+\exp(x){\rm Ei}(-x)\right]\right\}.
    \label{xlr3}
\end{equation}
The asymptotics of the short-range scaling function are obvious,
while these for the long-range one are
\begin{equation}
X^{{\rm lr}}(x)\simeq \left\{
\begin{array}{lc}
 24 \sqrt{2/\pi}\   x^{-4}\left(1+ O(x^{-6})\right),&
x \rightarrow \infty \\
& \\
\sqrt{2/\pi} \left(1+2 x^2\ln x\right )+
O(x^2),& x\rightarrow 0.
\end{array}
\right.
\label{cla3}
\end{equation}
\begin{figure}[tbp]
\epsfxsize=4.in \centerline{\epsffile{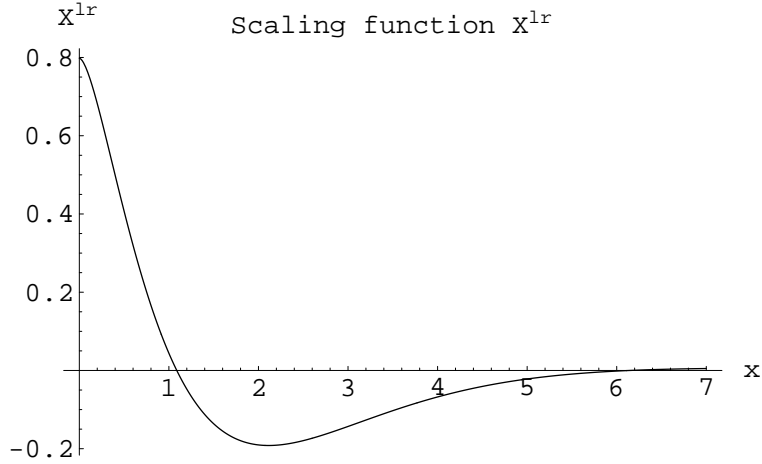}} \vspace{0.1in}
\caption{ The scaling function $X^{{\rm lr}}(x)$ of the long-range
correlation function for $d=\sigma=3$. One observes that, in
contrast with the short-range correlation function, it is {\it
not} a monotonic function of the scaling variable $x=r/\xi$. It
changes sign at $x \simeq 1.088$ and $x\simeq 6.146$ and reaches a
minimum at $x_{{\rm min}} \simeq 2.113$ which is $X^{{\rm
lr}}(x_{\rm min})\simeq -0.192$. In other words the long range
part of the interaction increases the correlations (in comparison
with an effective short-range system having the same value of $K$;
we recall that $K$ is a $\sigma$-dependent quantity) for $r$ up to
$1.088 \ \xi$ and for $r>6.145 \ \xi$, but decreases them for
$1.088 \ \xi<r<6.146 \ \xi$. The maximum of $X^{{\rm lr}}$ is
reached at $x=0$ and it is $X^{{\rm
    lr}}(0)=\sqrt{2/\pi}\simeq 0.798 $. The last implies, as it is to
be
expected, that for a fixed $r$  the maximal increment of the
correlations due to the long-range part of the interaction is reached
at $T=T_c$.  $X^{{\rm lr}}(x)$ decays in a
power law, as $x^{-4}$, for large values of its argument. }
\label{xlrfig}
\end{figure}

The asymptotics of the correlation function at $T=T_c$ and for any
fixed $\xi$ can be derived to much greater details in the limit
$r\rightarrow\infty$ for this especially important case. They are
(see Appendix \ref{A})
\begin{equation}
 G({\bf r};K|3,3)=\frac{b}{K}\frac{12}{\pi^2}  \frac{\xi^4}{r^6} \left[1+120
   (r/\xi)^{-2}+10080 (1-
\frac{2}{3}\frac{c}{\xi^2})(r/\xi)^{-4}+O((r/\xi)^{-6}) \right], \
r\gg\xi,
\label{cfa3}
\end{equation}
and
\begin{equation}
G({\bf r};K_c|3,3)=\frac{1}{K_c}\frac{1}{4\pi r}
 \left[1+\frac{2b}{\pi}r^{-1}-\frac{4}{\pi}b(b^2-2c)r^{-3}+O(r^{-5})
    \right], \ r\rightarrow\infty.
    \label{cfa3l}
\end{equation}
One can easily check that up to the leading-order terms these
asymptotics coincide with the corresponding ones that follow by
using the behavior of the short- and long-range correlation
functions given above. For example, when $r\rightarrow\infty$ but
$r\ll\xi$ (i.e. $x=r/\xi\rightarrow 0$ ) from Eqs. (\ref{xsr3})
and (\ref{xlr3}) one has
\begin{equation}
G({\bf r};K|3,3)=\frac{1}{K}\frac{1}{4\pi r}
 \left[1+x +\frac{1}{2}x^2 + \frac{2b}{\pi}r^{-1}-
\frac{4b}{\pi} r^{-1} x^2\ln{x}+O(r^{-3}, x^3, x^2 r^{-1})
    \right], \ r\rightarrow\infty,x=r/\xi\rightarrow 0.
\end{equation}

The crossover from short-range to long-range type behavior happens
at $r=r^*$ where $r^*$, in full agreement with Eq. (\ref{cover})
for $d=\sigma=3$, is given by
\begin{equation}
r^*=\xi\left\{\ln{\xi}+5 \ln\ln\xi + O\left(\frac{\ln\ln
      \xi}{\ln\xi}\right)
\right\}.
\end{equation}
Note that Eq. (\ref{cover}) was derived under the condition that
$d+\sigma<6$ and its ``analytical continuation'' to $d=\sigma=6$
is not obvious. It is nevertheless valid because for $r>>\xi$ the
leading-order term in the behavior of $G({\bf r};K|3,3)$ (see Eq.
(\ref{cfa3})) can be obtained from that one of $G({\bf
r};K|d,\sigma)$, given by Eq. (\ref{cfla}), if one sets
$d=\sigma=3$ there.
\begin{figure}[tbp]
\epsfxsize=4.in \centerline{\epsffile{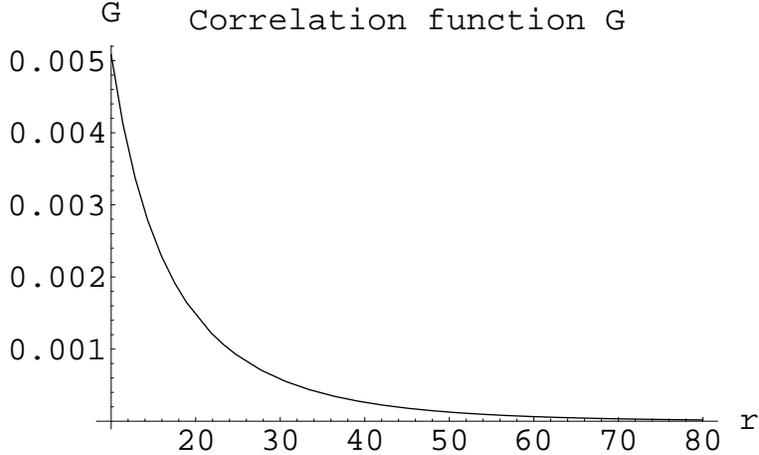}} \vspace{0.1in}
\caption{ We give here a representative example of the total
two-point correlation function $G({\bf r};K|d,\sigma)$ as a
function of $r$ for $d=\sigma=3$ and $K=b=1$, $\xi=20$. As it
should be expected for a ferromagnetic system $G>0$ and decays
monotonically as a function of $r$.
 }
\label{cfd3fig}
\end{figure}

\section{Finite-size scaling susceptibility}
\label{fsss}

One usually describes the critical behavior of finite systems in
the framework of the finite-size scaling theory
\cite{BDT00,F72,FB72,B83,P90}. The standard finite-size scaling is
usually formulated in terms of {\it only one} reference length,
namely the bulk correlation length $\xi$. The main statements of
the theory are that

{\it i)} The only relevant variable in terms of which the properties
of the finite system depend in the neighbourhood of the bulk critical
temperature $T_c$ is $L/\xi$.

{\it ii)} The rounding of the phase
transition in a given finite system sets in when $L/\xi=O(1)$.

The tacit assumption is that all other reference lengths (such as
lattice spacings, inverse cut-off, etc.) will lead {\it only to
corrections} in the above picture.  In addition, by analogy with
the bulk short-range systems it is supposed that if $\sigma\ge 2$
the finite-size critical behavior will be that of the
corresponding short-ranged finite-size systems (see, e.g.
\cite{FP86}), characterised  by exponentially fast decay of the
finite-size dependence of the thermodynamic quantities at least
when the critical region of the system is leaved in the direction
towards higher temperatures (the low-temperature behavior depends
on additional features like existence, or not, of spin-wave
excitations - Goldstone bosons).

As it has been recently shown the above picture is, in fact, more
complicated \cite{CD2000,DR2001,CD99} and not completely valid for
systems  with subleading long-range interactions \cite{DR2001}.
For such systems within  the mean-spherical model and under
periodic boundary conditions it has been found that the
finite-size susceptibility $\chi(t;L)$ is of the form
\cite{DR2001}
\begin{equation}
 \chi(t;L)=L^{\gamma/\nu}Y(x_1,bL^{2-\sigma}),
\end{equation}
or, equivalently,
\begin{equation}
 \chi(t;L)=L^{\gamma/\nu}
\left[Y^{{\rm sr}}(x_1) + bL^{2-\sigma} Y^{{\rm lr}}(x_1)\right],
\label{sl}
\end{equation}
where $x_1=atL^{1/\nu}$, and $Y$, $Y^{{\rm sr}}$ and $Y^{{\rm
lr}}$ are universal functions.  The quantities $a$,  and $b$ are
nonuniversal constants. One quite common way of fixing $a$ is to
choose it to be $a=(\xi_0^+)^{-1/\nu}$. It is worthily to note the
close similarity in the structures of Eqs. (\ref{cl}) and
(\ref{sl}). In other words - if one knows the structure of the
bulk two-point correlation function one easily can write the
corresponding finite-size behavior of the susceptibility. A
hypothesis about such a possibility has been stated for the first
time in \cite{CD2000}.

In the high-temperature, disordered phase, where
$tL^{1/\nu}\rightarrow\infty$, we find that the long-range portion
of the spin-spin interaction gives rise to contributions of the
order of $bL^{-(d+\sigma)}$ that swamp the exponentially small
terms that are expected to characterise the signature of finite
size in systems with periodic boundary conditions and short range
interactions.  In other words the {\it subleading} long-range part
of the interaction gives rise to a {\it dominant} finite-size
dependence in this regime.  This is entirely consistent with the
inherent long-range correlations that accompany long-range
interactions, but it violates the standard finite-size scaling
formulation. More explicitly, one obtains $Y^{{\rm sr}}(x_1)\sim
\exp(-{\rm const.} \ x_1^\nu)$, while
\begin{equation}
  Y^{{\rm lr}}(x_1)\sim x_1^{-d\nu-2\gamma},
\label{aY}
\end{equation}
when $x_1\rightarrow\infty$. This asymptotic follows from the
requirement the finite-size corrections to be of the order of
$L^{-(d+\sigma)}$ in this regime, which is to be expected on
general grounds and is supported by the existing both exact and
perturbative results for models with {\it leading} long-range
interaction included \cite{SP89,BD91,CT2000}. Note that (\ref{aY})
implies for the temperature dependence of this correction
\begin{equation}
 \chi(t;L)-\chi(t;\infty)\sim t^{-d\nu-2\gamma}L^{-(d+\sigma)},
\ tL^{1/\nu}\rightarrow\infty.
\end{equation}

Obviously, the existence of such power-law finite-size dependent
dominant terms above $T_c$ is of significance in the analysis of
Monte Carlo data for such systems.

Let us now consider the case $d+\sigma=6$, $2<d<4$, $2<\sigma<4$,
which contains the genuine van der Waals interaction with
$d=\sigma=3$. Then, instead of (\ref{sl}), one has \cite{DR2001}
\begin{equation}
 \chi(t;L)=L^{\gamma/\nu}
\left\{Y^{{\rm sr}}(x_1) + bL^{2-\sigma}
    \left[Y_1^{{\rm lr}}(x_1) \ln(L)+Y_2^{{\rm lr}}(x_1)
\right] \right\}.
\label{sl3}
\end{equation}
Comparing with the corresponding results for the correlation
function one observes, since there is no explicit $\ln r$
dependence there, that this subtle feature like the
logarithmic-in-$L$ corrections will not be captured in the above
mentioned approach --- from the structure of the bulk two-point
correlation function with respect to $r$ to obtain that one of the
finite size susceptibility with respect to $L$ (simply by
considering $L$ in the role of $r$). Nevertheless, indications
that the situation here may be more complicated are found in the
short distance expansion of the bulk correlation function which
has logarithmic in $r$ terms.

\section{Concluding Remarks and Discussion}
\label{d}

In the present article we derived the analytical behavior of the
two-point correlation function in a system with van der Waals type
interaction. The treatment has been made within the mean-spherical
model, which is an example of Ornstein and Zernike type theory. We
have pointed out that the leading order behavior of $G({\bf
r};T|d)$ as a function of the distance is exponential only within
the region of separations $r$ between the interacting objects
given by the condition $\xi \ll r \ll r^*\equiv (\sigma-2)\xi\ln
\xi$. Obviously, taking into account the dependence of $\xi$ on
the temperature, this region widens essentially only very close to
$T=T_c$. When $r$ is outside the region defined above the
correlations decay in a power law as a function of $r$: as
$r^{-(d-2)}$ for $r\ll \xi$, and as $r^{-(d+\sigma)}$ for $r\gg
(\sigma-2)\xi\ln \xi$. It turns out that $G({\bf r};T|d)$ can be
decomposed in a "short-range" and "long-range" parts (see Eq.
(\ref{cs})). The corresponding short- and long-range scaling
functions $X^{\rm sr}$ and $X^{\rm lr}$ are given in Eqs.
(\ref{cse}) and (\ref{cle}), respectively. For the case $2<d<4$,
$2<\sigma<4$ and $d+\sigma<6$ the behavior of these functions is
illustrated in Fig. \ref{xsr_xlr_fig}, whereas the small and large
value asymptotics of the functions are given in Eqs. (\ref{csa})
and (\ref{cla}). A special attention is paid to the most important
case of $d=\sigma=3$ which mimics the real van der Waals
interaction in fluids. The analytical expressions for the scaling
functions are given in Eqs. (\ref{xsr3}) and (\ref{xlr3}). The
behavior of $X^{\rm lr}$ is plotted on Fig. \ref{xlrfig}. The
asymptotics of $G({\bf r};K|3,3)$ and $G({\bf r};K_c|3,3)$ are
given in Eqs. (\ref{cfa3}) and (\ref{cfa3l}). The behavior of the
total correlation function $G({\bf r};K|3,3)$ is illustrated in
Fig. \ref{cfd3fig}.

Let us note that since all of the above results are for the mean
spherical model they pertain to the case of $\eta=0$ models.
Naturally, one stacks with the question: How expressions like
(\ref{cl}) and (\ref{cover}) should be modified for models with
$\eta \ne 0$? A hint in this direction can be found in \cite{KR84}
- for such models Kayser and Ravech\'{e} suggest that, in our
terminology, $r^*=(\sigma-2+\eta)\xi \ln \xi$. In order to
reconcile this statement with Eq. (\ref{cl}) one has to suppose
that when $\eta\ne0$ (see also Eq. (\ref{cor}))
\begin{equation}
G(r;K|d,\sigma) = D(T)r^{-(d-2+\eta)} \left[X^{\rm{\pm,
sr}}(r/\xi) +br^{-(\sigma-2+\eta)}X^{\rm{\pm, lr}}(r/\xi) +\cdots
\right]. \label{clmod}
\end{equation}
Here $X^{\rm sr}$ is supposed to have the usual properties (see
Eq. (\ref{cplus})), whereas for $X^{\rm lr}$ we suppose that
$X^{\rm lr}(x)\rightarrow X^{\rm lr}_-$, $x\rightarrow 0$ and
$X^{\rm lr}(x)\rightarrow X^{\rm lr}_+ x^{-2(2-\eta)}$,
$x\rightarrow \infty$, where $X^{\rm lr}_-$ and $X^{\rm lr}_+$ are
positive constants. The large value asymptotics of $X^{\rm lr}(x)$
ensures that the correlation function decays as $r^{-(d+\sigma)}$
for $r\gg r^*$, which is in full agreement with (\ref{theorem}),
and that $r^*=(\sigma-2+\eta)\xi \ln \xi$, which coincides with
the result of Kayser and Ravech\'{e} \cite{KR84}. The property
$G({\bf r};T|d) \simeq J(r) \chi^2/\beta$, $r\rightarrow\infty$
\cite{KR84} is retained too.

We emphasize, nevertheless, that despite all of the above
features, for the moment (\ref{clmod}) is only a plausible
hypothesis the verification of which is still lacking.

At the very end we note that, according to a recent hypothesis
\cite{CD2000},  the behavior of the {\it bulk} two-point
correlation function $G(r;K|d,\sigma)$ can be related to that one
of the {\it finite-size} susceptibility under periodic boundary
conditions. In the present article (see Section \ref{fsss}) we
checked this hypothesis within the mean spherical model using the
results for the finite-size susceptibility derived in
\cite{DR2001}. Definitely, one can extend the calculations
presented here to models with $\eta\ne0$ by using renormalization
group techniques.

\section{Acknowledgements}

Support by DAAD, by DLR under grant number 50 WM 9911 and by NASA
under contract number 1201186 is gratefully acknowledged.

The author thanks Prof. V. Dohm for the stimulating discussions,
and Prof. J. Brankov, S. Dietrich, J. Rudnick, N. Tonchev, Dr. M.
Krech and Dr. E. Korutcheva  for the exchange of information and
critical reading of the manuscript. The author is also indebted to
Prof. P. Russev for drawing his attention to Ref. {\cite{L75}}
concerning the properties of the generalized hypergeometric
functions.

\appendix
\section{MATHEMATICAL DETAILS}
%\section{Large value asymptotic behavior of the generalized
 % hypergeometric function $_1 F_2(a;b,c;z)$}
\label{A}

In this Section we will provide the mathematical details needed to
derive Eqs. (\ref{cla}), (\ref{cla3}), (\ref{cfa3}) and
(\ref{cfa3l}).

Let us start with the case $d=\sigma=3$. Then
\begin{equation}\label{G3}
  G({\bf
  r};K|3,3)=\frac{1}{K}\frac{1}{2\pi^2}\frac{1}{r}\int_0^\infty
  f(k)\sin(k r)dk,
\end{equation}
where
\begin{equation}\label{f}
  f(k):=\frac{k}{\xi^{-2}+k^2-bk^3+ck^4}.
\end{equation}
It is easy to show that if for a given integer $n>0$ the
derivatives $f^{(p)}(0)$ of the function $f$ do exist for
$p=0,\cdots, 4n+2$ and, in addition, $f^{(p)}(\infty)=0$,
$p=0,\cdots, 4n+2$, then
\begin{equation}\label{fp}
\int_0^\infty f(k)\sin(k r)dk=r^{-1}\sum_{p=0}^n
\left[f^{(4p)}(0)r^{-4p}-f^{(4p+2)}(0)r^{-4p-2}
\right]+O(r^{-4p-5}).
\end{equation}
Applying this to (\ref{G3}) in the limit $r\gg\xi$ one immediately
obtains (\ref{cfa3}).

In order to derive Eq. (\ref{cfa3l}) let us note that
($\xi^{-2}=0$ at $K=K_c$)
\begin{eqnarray}
  G({\bf
  r};K_c|3,3)&=&\frac{1}{K_c}\frac{1}{2\pi^2}\frac{1}{r}\int_0^\infty
  \frac{1}{1-bk+ck^2}\frac{\sin(k r)}{k} dk  \nonumber \\
  &=&\frac{1}{K_c}\frac{1}{2\pi^2}\frac{1}{r}\left[ \frac{\pi}{2}+
  b \int_0^\infty f_b(k)\sin(k r)dk-c \int_0^\infty f_c(k)\sin(k r)dk
  \right],
\label{G3Kc}
\end{eqnarray}
where
\begin{equation}\label{fbd}
  f_b(k):=\frac{1}{1-b k+c k^2},
\end{equation}
and
\begin{equation}\label{fcd}
  f_c(k):=\frac{k}{1-b k+c k^2},
\end{equation}
and use has been made of the fact that $\int_0^\infty
\sin(k)/k=\pi/2$. Applying again (\ref{fp}) for the evaluation of
the integrals in Eq. (\ref{G3Kc}), we obtain the result given in
Eq. (\ref{cfa3l}).

The derivation of Eq. (\ref{cla3}) is a bit more complicated.
First, let us note that
\begin{eqnarray}
G({\bf r};K|3,3)&=&\frac{K_c}{K}G({\bf
r};K_c|3,3)+\frac{1}{K}\frac{1}{2\pi^2}\frac{1}{r}\int_0^\infty
  \left[\frac{k^2}{\xi^{-2}+k^2-bk^3+ck^4}-
  \frac{k^2}{k^2-bk^3+ck^4}\right]\frac{\sin(k r)}{k} dk  \nonumber \\
  &=&\frac{K_c}{K}G({\bf
r};K_c|3,3)-\frac{1}{K}\frac{1}{2\pi^2}\frac{\xi^{-2}}{r}\int_0^\infty
\frac{1}{(1-bk+ck^2)(\xi^{-2}+k^2-bk^3+ck^4)} \frac{\sin(k r)}{k}
dk \nonumber \\
&\simeq& \frac{K_c}{K}G({\bf
r};K_c|3,3)-\frac{1}{K}\frac{1}{2\pi^2}\frac{\xi^{-2}}{r}\left[\int_0^\infty
\frac{1}{(1-bk+ck^2)(\xi^{-2}+k^2)} \frac{\sin(k r)}{k} dk
+\right. \nonumber \\
&& \left. b \int_0^\infty \frac{k^3}{(\xi^{-2}+k^2)^2}
\frac{\sin(k r)}{k} dk \right] \nonumber \\
&\simeq &\frac{K_c}{K}G({\bf
r};K_c|3,3)-\frac{1}{K}\frac{1}{2\pi^2}\frac{\xi^{-2}}{r}\left[\int_0^\infty
\frac{1}{\xi^{-2}+k^2} \frac{\sin(k r)}{k} dk +b \int_0^\infty
\frac{1}{\xi^{-2}+k^2} \sin(k r) dk
+\right. \nonumber \\
&& \left. b \int_0^\infty \frac{k^2}{(\xi^{-2}+k^2)^2}
\frac{\sin(k r)}{k} dk \right]\nonumber \\
&=& \frac{K_c}{K}G({\bf
r};K_c|3,3)-\frac{1}{K}\frac{1}{2\pi^2}\frac{\xi^{-2}}{r}\left[\int_0^\infty
\frac{1}{\xi^{-2}+k^2} \frac{\sin(k r)}{k} dk +b
(2+\xi^{-2}\frac{\partial}{\partial \xi^{-2}} )\int_0^\infty
\frac{1}{\xi^{-2}+k^2} \sin(k r) dk \right].
\end{eqnarray}
Above we have already dealt with the large distance asymptotic of
$G({\bf r};K_c|3,3)$. In order to obtain (\ref{cla3}) now it only
remains to note that \cite{RG73}
\begin{equation}\label{int1}
   \xi^{-2}\int_0^\infty
\frac{1}{\xi^{-2}+k^2} \frac{\sin(k r)}{k}
dk=\frac{\pi}{2}\left[1-\exp\left(-\frac{r}{\xi}\right)\right],
\end{equation}
and
\begin{equation}\label{int22}
  \int_0^\infty
\frac{\sin(k)}{x^{2}+k^2} dk=\frac{1}{2x}\left[ \exp(-x){\rm
Ei}(x)-\exp(x){\rm Ei}(-x)\right], \ {\rm Re}(x)>0.
\end{equation}

At the end, let us derive the results given in Eq. (\ref{cla}).
The case $1 \ll r\ll\xi$, i.e. $x\rightarrow 0$, is simple - using
the series representations of the modified Bessel functions
$I_a(x)$ (see, e.g. \cite{AS70} or \cite{RG73}) one obtains from
Eqs. (\ref{cle})-(\ref{ghf}) the asymptotics of $X^{{\rm lr}}(x)$
for small values of the argument, given in Eq. (\ref{cla}). Much
more interesting is the case when $r\gg\xi\gg 1$, i.e. when $x\gg
1$. We will present here a derivation of the corresponding
asymptotic of $X^{{\rm lr}}(x)$ without making use of the large
value asymptotic results for the function $_1F_2$. One can get an
impression of the beauty of the proposed way of acting only after
taking a look at the results available for the function $_1F_2$
(see, e.g., \cite{L75}). We start by noting that
\begin{eqnarray}
    X^{\rm lr}(x) &=& \int_0^\infty dt \ \frac{t^{\sigma+d/2}}{(x+t^2)^2}J_{d/2-1}(t)
    \nonumber \\
     &=& (1+x\frac{\partial}{\partial x})\int_0^\infty dt \
     \frac{t^{\sigma+d/2}}{x+t^2}J_{d/2-1}(t) \nonumber \\
     &=&(1+x\frac{\partial}{\partial x})\int_0^\infty dz \ \exp(-z
     x)\int_0^\infty dt \ \exp(-z t^2)t^{\sigma+d/2-2}J_{d/2-1}(t).
     \label{lrad}
\end{eqnarray}
In order to evaluate the last integral in (\ref{lrad}) one can use
the formula \cite{RG73}
\begin{equation}\label{RGh}
  \int_0^\infty dx \ x^\mu \exp(-\alpha x^2) J_\nu (\beta x)=
  \frac{\beta^\nu
  \Gamma(\nu/2+\mu/2+1/2)}{2^{\nu+1}\alpha^{(\mu+\nu+1)/2}\Gamma(\nu+1)}
  \
  _1F_1\left(\frac{\nu+\mu+1}{2};\nu+1;-\frac{\beta^2}{4\alpha}
  \right).
\end{equation}
With its help one obtains
\begin{equation}\label{d1f1}
\int_0^\infty dt \ \exp(-z
t^2)t^{\sigma+d/2-2}J_{d/2-1}(t)=\frac{\Gamma\left(\frac{d+\sigma}{2}-1\right)}{2^{d/2}
\Gamma\left(\frac{d}{2}\right)}z^{1-(d+\sigma)/2} \ _1F_1\left(
\frac{d+\sigma}{2}-1;\frac{d}{2};-\frac{1}{4z} \right),
\end{equation}
i.e.
\begin{equation}\label{gexpr}
X^{\rm
lr}(x)=\frac{\Gamma\left(\frac{d+\sigma}{2}-1\right)}{2^{d/2}
\Gamma\left(\frac{d}{2}\right)}(1+x\frac{\partial}{\partial
x})\int_0^\infty dz \ \exp(-z x) z^{1-(d+\sigma)/2} \ _1F_1\left(
\frac{d+\sigma}{2}-1;\frac{d}{2};-\frac{1}{4z} \right).
\end{equation}
Note now that when $x\gg1$ the main contribution of the integral
in the above expression will stem from small $z$ values. Using the
corresponding asymptotic \cite{AS70}
\begin{equation}\label{asf11}
  _1F_1(a;b;-y)=\frac{\Gamma(b)}{\Gamma(b-a)}y^{-a}(1+a(a-b+1)y^{-1}+O(y^{-2}))
\end{equation}
of $_1F_1(a;b,-y)$ for $y\gg 1$ and performing the integrations,
we arrive at the asymptotic of $X^{\rm lr}(x)$ reported in Eq.
(\ref{cla}) for the case $x\gg1$.


\begin{references}
\bibitem{note2} In order to write the above equations in the
form given we have taken the number density of the fluid to be $\rho=1$.
\end{references}
\end{document}